    \documentstyle[proceedings]{crckapb} 

\begin{opening}
\title{HIGH PRECISION ASTROMETRY WITH VLBI:
       \protect\\
FROM THE TRIANGLE 1803+784/1928+738/2007+777 
       \protect\\
TO THE COMPLETE S5 POLAR CAP SAMPLE}
\author{E. ROS}
\institute{Max-Planck-Institut f\"ur Radioastronomie, 
           Bonn, Germany}
\author{J.M. MARCAIDE}
\author{J.C. GUIRADO}
\author{M.A. P\'EREZ-TORRES}
\institute{Dept.\ d'Astronomia i Astrof\'{\i}sica,
           U.\
           de Val\`encia, 
           Spain}

\end{opening}

\runningtitle{HIGH PRECISION ASTROMETRY WITH VLBI}

\begin{document}



\section{Introduction}
The Very Long Baseline Interferometry (VLBI) technique can image 
compact radio sources with a resolution of the order of the
milliarcsecond and can determine astrometrically relative positions
to precisions of the order of tens of microarcseconds.  
This technique is ideal to construct a precise
celestial reference
frame.  Up to the present, the group delay observable has been 
regularly used in such a task. 
The use of the more precise phase delay observable should constitute
an immediate improvement in accuracy.
Furthermore, the phase-delay, if
used differentially over radio source pairs, becomes the
most accurate observable in astrometry.  
The technique of phase-delay differential astrometry has been applied
to several source pairs, with separations ranging from 33$^{\prime \prime}$  
(1038+528 A/B, \citeauthor{mar83}, \citeyear{mar83})
to 5$\rlap{.}^\circ$9 (3C\,395/3C\,382, \citeauthor{lar96}, 
\citeyear{lar96}).
The pair 1928+738/2007+777 
(4$\rlap{.}^\circ$6 separation) has been also studied (\citeauthor{gui95}
\citeyear{gui95,gui98})
yielding precisions of $\sim$200\,$\mu$as.  Recently,
we added a new source 1803+784 to the latter
pair to take advantage of the constraints
introduced by a triangular geometry in the determination of the angular
separations \cite{ros98}.  It represents a first step towards extending
the differential
phase-delay astrometry from pairs to a whole sky radio source frame.

\section{The Triplet 1803+784/1928+738/2007+777.}
We observed the radio sources of the triangle formed by the BL-Lac
objects 1803+784 and 2007+777, and the QSO 1928+738 
on epoch
1991.89 with an intercontinental interferometric array simultaneously
at the frequencies of 2.3 and 8.4\,GHz \cite{ros98}.  
We determined the angular separations
among the three radio sources with submilliarcsecond accuracy from a weighted
least squares analysis to the differential phases, after removing
most of the contribution due to the geometry of the array and the 
atmosphere.
The radio source structure contributions to the phase delays
were also modeled using hybrid mapping images of the radio sources
from the same observations.  
We checked the consistency of our astrometric determination through the
use of the so-called
Sky-Closure.  The Sky-Closure was defined as the circular 
sum of the angular separations of the
three radio sources, determined pairwise and independently.  In our
case the result was consistent with zero, and verified satisfactorily
the data process followed.  The final accuracy of the astrometric 
determinations was of 130\,$\mu$as.

\begin{figure}[htb]
\vspace{76mm}
\includegraphics{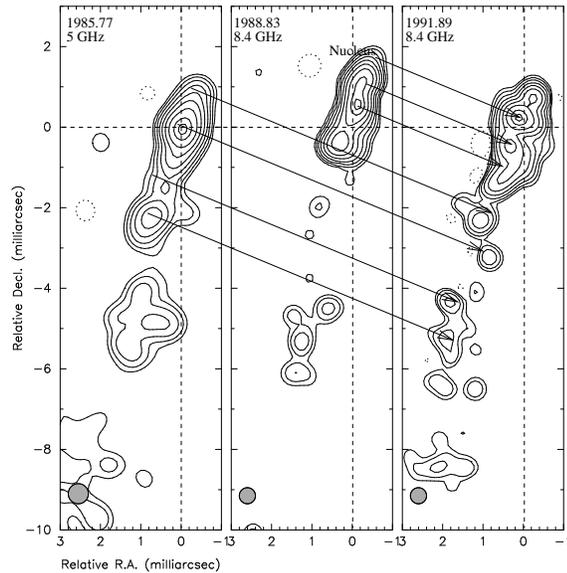}
\caption{Astrometric alignment of the maps of 1928+738.  We infer a proper 
motion of the components of about 0.3\,mas/yr towards the south, and also 
that the nucleus of the radio source is near, but north of, the northernmost 
component of the 1988.83 map, at the position indicated in the central figure.
The components have been convolved with circular gaussian beams (bottom,
left) of diameters 0.7, 0.4, and 0.4\,mas, showing slightly overresolved
images.}
\label{fig:19motions}
\end{figure}

One important aspect in the astrometric work is the excess
propagation delay due to the ionization of the propagation medium,
mainly the ionosphere.  
The ionospheric
contribution to the delays had been
determined in the past from dual-band VLBI
observations.  
\citeauthor{sar94} \shortcite{sar94} showed
that the total electron content (TEC) of the ionosphere can be determined
with high accuracy by using dual
frequency Global Positioning System (GPS) data.  
We used their method to estimate the plasma contribution by
using TEC estimates of the ionosphere
obtained from data from different GPS sites neighbor to the VLBI
stations (see \citeauthor{ros98} \citeyear{ros98,ros99}).  
\paragraph{Proper Motions in 1928+738.}
The comparison of the measurements of the separations of the pair 
1928+738/2007+777 with those presented by \citeauthor{gui95} 
\shortcite{gui95,gui98} for epochs 1985.77 and 1988.83
allows us to register adequately the
absolute position of 1928+738 relative to 2007+777.  We estimate
the proper motion of components in 1928+738, and identify the
position of the radio source core even though it is unseen at cm-wavelengths, 
as shown in Fig.\ \ref{fig:19motions}.
The average proper motion of the components emerging from the core
region is of 0.30$\pm$0.15\,mas/yr towards the South.

\section{The S5 Polar Cap Sample.}
\begin{figure}[htb]
\vspace{40mm}
\includegraphics{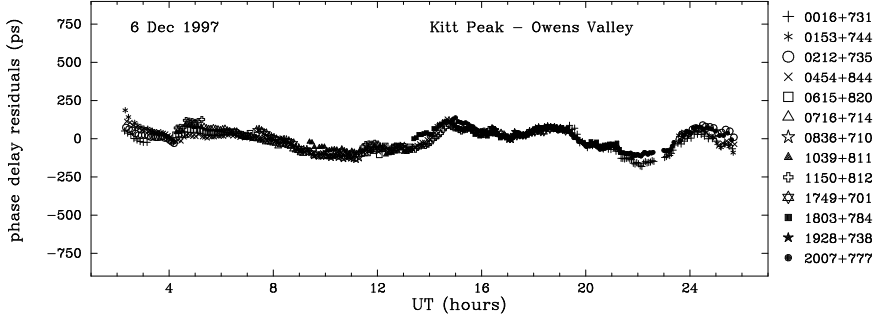}
\caption{Preliminary results of the first epoch of 8.4\,GHz observations of
the complete S5 polar cap sample:
postfit residuals of the (undifferenced) phase-delays after a weighted
least-squares analysis which estimates the relative separations among the
radio sources. One phase-cycle at the frequency of 8.4\,GHz corresponds
to  120\,ps of phase-delay. }
\label{fig:phascon}
\end{figure}

The phase connection over a separation as large as 15$^\circ$ is critical
for the success of the extension of the astrometry to larger samples.
\citeauthor{per99} \shortcite{per99} have shown the phase-connection
for such a pair, 1150+812/1803+784, to be possible.  
Therefore, the 13 radio sources of the complete S5 polar cap sample
\cite{eck86}, which have mutual separations less than 15$^\circ$ can
be studied also astrometrically.  
These sources are the quasars
0016+731, 0153+744, 0212+735, 0615+820, 0836+710, 1039+811,
1150+812, and 1928+738, and the BL Lacertae objects 0454+844,
0716+714, 1749+701, 1803+784 and 2007+777.  
We observed this
set of radio sources at 8.4\,GHz over 24 hours on epoch 1997.93.  We
imaged the 13 radio sources using hybrid mapping techniques.
On these
images we defined reference points and then removed the structure
contributions from the corresponding
astrometric observables.  After it,
we used the differential phase-delays to
obtain a global solution of all the source positions.  
Fig.\ \ref{fig:phascon}
shows our preliminary results.  A similar trend of systematic
effects, which will cancel out when making the differences, is conspicuous.

With respect to the determination of the ionosphere contribution to
the data, the density of the GPS network increased notably from 1991 to 1997,
making the bias removal and the accuracy of the TEC
determination much better.  
Now it is possible to have Global Ionospheric Maps from
the Global Positioning System
and thus 
estimate the 
ionosphere contribution to the astrometric observables of a single-wavelength
VLBI observation and to remove the plasma effects from them with high accuracy.

\section{Conclusions}
The differential phase-delay astrometry has recently undergone important 
improvements.  The phase-connection process has been
extended to larger sets of radio sources with larger source 
separations, and the 
ionosphere contribution to the astrometric observables has been
successfully removed using GPS
data.  We have determined with submilliarcsecond precision the
relative separations in the triangle of radio sources 
1803+784/ 1928+738/ 2007+777, and we have observed astrometrically 
the complete S5 polar cap sample, that among the 13 sources
within  20$^\circ$ to the celestial North Pole
includes the above 3. 
New observations
are now underway in the framework of a long-term astrometric program
to determine the absolute kinematics of radio source components in
the S5 complete sample.  This program, extended
over 5 years, will reach a precision in the determination of the
relative separations better than 0.1\,mas and consequently in the 
proper motions of the radio source components. 

%

\end{document}